\documentclass[twocolumn,aps,prb,longbibliography,superscriptaddress]{revtex4}
\usepackage{graphics,bm,graphicx}
\usepackage{amssymb}
\usepackage{epsfig}
\usepackage{epsf}
\usepackage{xcolor}
\usepackage{colortbl}
\usepackage{natbib}
\usepackage[colorlinks=true,allcolors=blue]{hyperref}
\usepackage{hypernat} 

\def\avg#1{\langle#1\rangle}

\def\be{\begin{equation}} \def\ee{\end{equation}}
\def\beq{\begin{eqnarray}} \def\eeq{\end{eqnarray}}

\def\nn{\nonumber}

\begin{document}

\title{Stripe spin-density-wave and chiral superconductivity in tWSe$_2$
}

\author{Erekle Jmukhadze}
\affiliation{Department of Physics, Applied Physics, and Astronomy, Binghamton University, Binghamton, New York, 13902, USA}

\author{Sam Olin}
\affiliation{Department of Physics, Applied Physics, and Astronomy, Binghamton University, Binghamton, New York, 13902, USA}

\author{Allan H. MacDonald}
\affiliation{Department of Physics, the University of Texas at Austin, Austin, Texas 78712, USA}

\author{Wei-Cheng Lee}
\email{wlee@binghamton.edu}
\affiliation{Department of Physics, Applied Physics, and Astronomy, Binghamton University, Binghamton, New York, 13902, USA}

\date{\today}

\begin{abstract}
The layer-dependent Hamiltonians of parallel-stacked MoTe$_2$ and WSe$_2$ homobilayer moir\'e materials are topologically non-trivial, both in real space and in momentum space, and have been shown to support integer and fractional quantum anomalous Hall states, as well as antiferromagnetic and superconducting states. 
Here, we address the interplay between the antiferromagnetic and superconducting states observed in tWSe$_2$ when the Fermi level is close to its $M$-point van Hove singularity
and the displacement field is small.  
We combine DFT with path-integrals to construct a minimal
moir\'e band model that accounts for lattice relaxation along the $c$-axis and   
perform Hartree-Fock calculations to identify competing charge and spin ordered states. 
For tWSe$_2$ at $\theta=2.7^\circ$ and $\theta=3.65^\circ$,
we find that a layer antiferromagnet (AFM), a stripe spin-density-wave (SDW),
and the ferromagnetic Chern insulator (FM) are the primary candidates for the ground state at zero displacement field, and 
argue that antiferromagnetic spin interactions on the next neighbor bond 
$J_2$ can induce a time-reversal symmetry breaking chiral superconducting state.
\end{abstract}

\maketitle

\section{Introduction}
Over the past decade, experimental and theoretical advances have 
demonstrated the richness of two-dimensional van der Waals moir\'{e} materials [\onlinecite{Bistritzer_2011,wu2019topological,andrei2021marvels}] as a platform for tunable strongly correlated states, including, for example, Mott insulators [\onlinecite{PhysRevLett.121.026402,Li2021}], Kondo lattice states [\onlinecite{PhysRevB.106.L041116,PhysRevLett.129.047601,Zhao2023}], unconventional magnets with tunable
exchange interactions [\onlinecite{Tang2023,doi:10.1126/science.adg4268,Ciorciaro2023}], and room-temperature
memristors [\onlinecite{Yan2023}]. Recently, observations of fractional Chern insulator (FCI) states in tMoTe$_2$ [\onlinecite{Zeng2023,Cai2023,Park2023,PhysRevX.13.031037}] homobilayers
and rhombohedral pentalayer-graphene/hBN moir\'{e} materials [\onlinecite{Lu2024}], and superconductivity (SC) in both tWSe$_2$ [\onlinecite{Xia2024, guo2024, dean2026, xia2026}] and
tMoTe$_2$ [\onlinecite{fanxu2025}] homobilayers
have promoted moir\'{e} physics to a new level of significance.  
In both graphene and TMD systems, many questions about 
the FCI states remain, in particular in 
explaining the moir\'{e} band filling factors and twist angles at which they appear and differences between materials. For example, while FCIs were observed experimentally in twisted homobilayer MoTe$_2$ (tMoTe$_2$) 
at the hole filling of $\nu=-\frac{2}{3}$, no broken symmetries were observed at the 
particle-hole conjugate filling factor $\nu=-\frac{4}{3}$, in
disagreement with theoretical predictions [\onlinecite{PhysRevLett.132.036501}]. 
Another important challenge is to understand the nature of the 
superconducting states observed in both tWSe$_2$ [\onlinecite{Xia2024, guo2024, dean2026, xia2026}] and  tMoTe$_2$ [\onlinecite{fanxu2025}], whose non-superconducting phase diagrams appear to be very different. 

The emergence of a particular order in a system
is subject to a delicate interplay between 
band-topology and electronic correlations  [\onlinecite{PhysRevLett.106.236803,PhysRevLett.106.236804,PhysRevLett.106.236802,Sheng2011,PhysRevX.1.021014,PhysRevB.85.075116, Dan1}]. 
To shed light on when one state is most likely to be preferred over others in TMD homobilayer moir\'e materials,
we perform full moir\'e continuum model mean-field calculations, concentrating on the case of a
moir\'e band hole filling factor $\nu=1$.  
We perform our calculations using a continuum model [\onlinecite{PhysRevLett.118.147401}] 
constructed from density-functional theory (DFT) calculations 
that account for local structural relaxation along the $c$-axis.
The model is derived by building accurate Wannier tight-binding models at all stacking arrangements 
and integrating out bands that are far from the Fermi energy. 
Compared to direct large-scale DFT calculations [\onlinecite{PhysRevLett.132.036501,PhysRevB.109.205121,Zhang2024}], this approach is applicable over a wider range of twist angles and has lower computational cost. 
The moir\'{e} model is constructed directly and does not rely on additional fits that assume smooth spatial variations.
We do not account for in-plane strain relaxation, since we are interested in intermediate twist-angle 
properties.

In this paper, we construct moir\'{e} models for 
tWSe$_2$ and tMoTe$_2$.  We then focus on the phase diagram of tWSe$_2$, which is less studied compared to tMoTe$_2$  [\onlinecite{PhysRevX.13.041026,PhysRevLett.133.026801}] and 
has a distinct phase diagram.  We first perform Hartree-Fock calculations on the moir\'{e} model using gate-screened Coulomb potentials to identify the competing spin and charge orders.  In addition to the multiferroic states with inter-layer ferroelectricity (FE) and  ferromagnetism (FM) identified in previous MoTe$_2$ studies, 
we find that at zero displacement field 
and for twist angles between $\theta=2.7^\circ$ and $\theta=3.65^\circ$,
tWSe$_2$ can support a competing stripe spin-density-wave (SDW) state whose ordering wavevector is associated with $M$-point van Hove singularities in the moir\'{e}
band structure.  A metallic layer AFM state exists
with a mean-field energy that is only slightly higher than that of the stripe SDW state. Based on our analysis of the antiferromagnetic ordering patterns in these low energy states of tWSe$_2$, and the strong onsite Hubbard repulsion in the narrow moir\'{e} bands, we are able to construct a $t$–$J$–$U$ model that provides a possible explanation for the origin
of superconductivity in tWSe$_2$ in the framework of a two-band model that is known to 
qualitatively capture the physics.  We conclude that second neighbor  
superexchange $J_2$ can induce intra-layer Cooper pairs.  
The resulting superconducting phase is a chiral state that spontaneously breaks time-reversal symmetry and has 
a mixture of singlet and triplet components. Our results suggest that competition between the stripe SDW state and superconductivity underlies the first-order transition observed at small displacement fields in this system.

\section{Model construction}
\begin{figure}
\begin{center}
\includegraphics[width=3.4in]{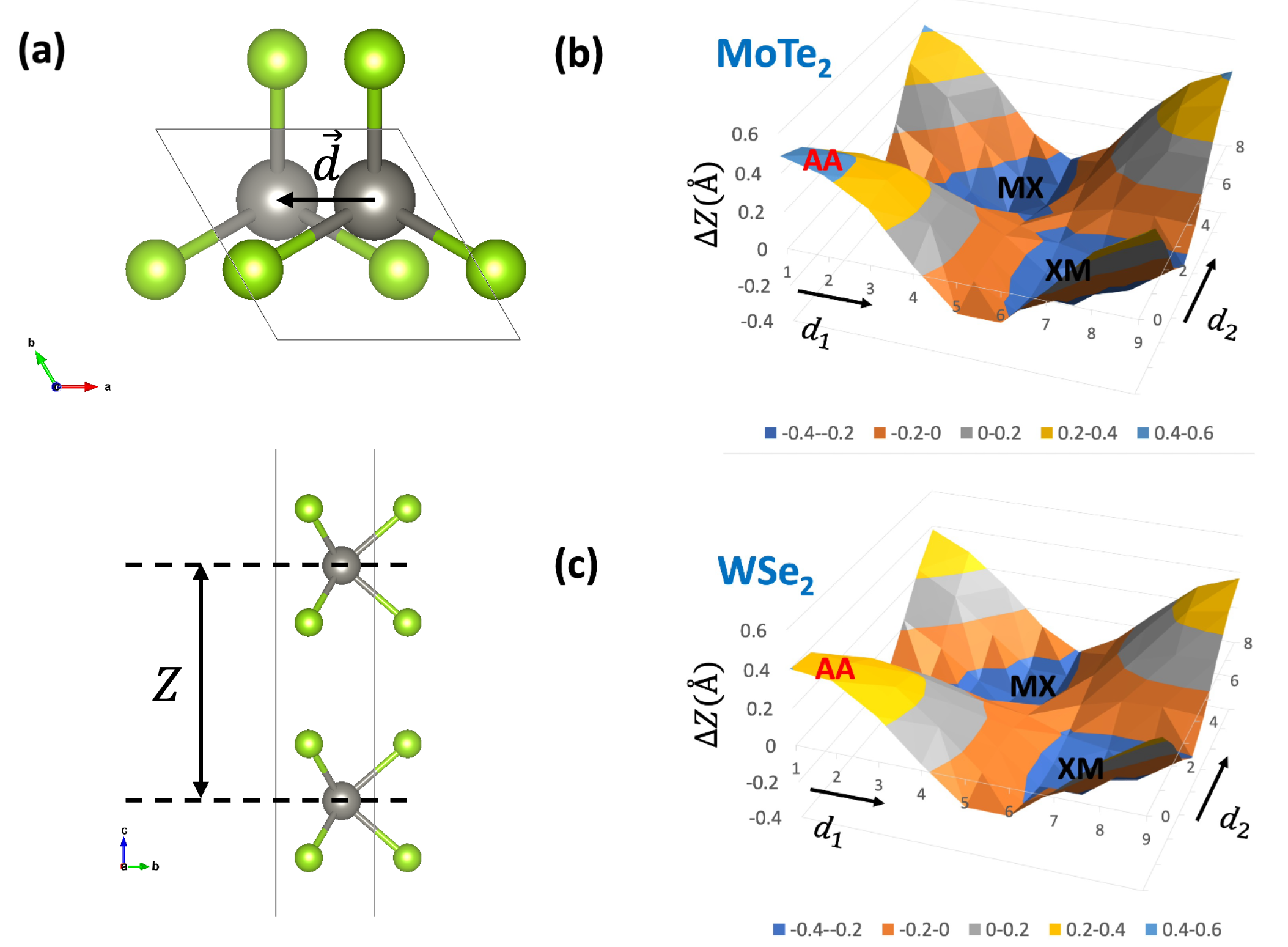}
\caption{\label{fig:displacement} (a) Schematic illustration of the displacement $\vec{d}$ and the metal atom
separation $Z$ along the $c$-axis. (b) Variation of $Z$ across the $\vec{d}$-grid after $c$-axis relaxation for homobilayer MoTe$_2$.  ($\vec{d}$ is defined to be zero at AA stacking.)
$\Delta Z(\vec{d})\equiv Z(\vec{d}) - Z_m$, where $Z_m=7.279$\r{A} is the spatial average $Z(\vec{d})$. The same plot for WSe$_2$ for which $Z_m=6.726$\r{A}.  The layer separations are minimized at the equilibrium 
metal on chalcogen (MX) and chalcogen on metal (XM) stacking points.} 
\end{center}
\end{figure}

Following the recipe outlined in Ref. [\onlinecite{PhysRevLett.118.147401}], we specify stacking by fixing the shift $\vec{d}$ in the lateral projection of the top layer metal atom relative to its bottom layer counterpart, as shown in Fig.~\ref{fig:displacement}(a). At a given $\vec{d}$, the metal ions are allowed to move freely only along the $c$-axis, while all the chalcogenide ions can move freely in all directions.  We vary $\vec{d}$ over one unit cell of the untwisted lattice: 
$\vec{d}=d_1 \vec{a}_1+d_2\vec{a}_2$, where $0\leq d_1,d_2 < 1$, and $\vec{a}_1 = a_0(1,0,0)$ and
$\vec{a}_2 = a_0(-\frac{1}{2},\frac{\sqrt{3}}{2},0)$ are the unit vectors of the untwisted lattice.
($a_0$ is the in-plane lattice constant.) In our calculations, we chose a $9\times 9$ grid for $(d_1,d_2)$. 
The Vienna \emph{ab initio} Simulation Package (VASP) [\onlinecite{vasp1,vasp2,vasp3}] is employed to perform all the structural relaxation and band structure calculations. The structural relaxation parameters are the same as in our previous work [\onlinecite{PhysRevB.109.165101}], and the LDA functional with spin-orbit coupling (LDA+SO) is adopted for all the first-principles calculations. 
After completion of structural relaxation, we construct a tight-binding (TB) model 
including the $d$ orbitals of the metal ions (Mo, W) and the 
$p$ orbitals of the chalcogenide ions (Te, Se) using 
WANNIER90 [\onlinecite{wannier90}] for each displacement $\vec{d}$. 
Since the dependence on $\vec{d}$   is periodic,
the dependence of the TB Hamiltonian on $\vec{d}$ reflects the lattice periodicity 
and has a corresponding Fourier representation. 

Now we are ready to derive the bilayer moir\'e material continuum model.
Due to strong spin-orbit coupling, the valence band maximum (VBM) at $\vec{K}$  ($\vec{K}'$) is strongly spin-polarized, and the orbital angular moment tends to align with the spin moment. This suggests that the $Y_2^{\pm 2}$
d-orbital (+ at $K$ and - at $K'$) has the
largest weight at the $\vec{K}$ ($\vec{K}'$) VBM,
as has been confirmed in many first-principles calculations. 
As a result, our goal is to obtain an effective two-level model in the basis of
$Y_2^{\pm 2}$ orbitals at the metal sites in opposite layers.
To achieve this goal, we first rewrite the TB model at displacement $\vec{d}$ as
\beq
\hat{H}(\vec{k},\vec{d}) &=& \hat{h} + \hat{H}_R + \hat{W},\nn\\
\hat{h}&=&\sum_{i,j=t,b} h_{ij}\hat{c}^\dagger_i\hat{c}_j = \phi^\dagger {\bf h} \phi,\nn\\
\hat{H}^R&=& \sum_{m,n\neq t,b} H^R_{mn}\hat{c}_m^\dagger \hat{c}_n=\chi^\dagger {\bf H^R}\chi,\nn\\
\hat{W}&=&\sum_{i=t,b}\sum_{m\neq t,b} W_{mi}\hat{c}_m^\dagger\hat{c}_i + h.c.
= \chi^\dagger {\bf W} \phi
\eeq
where $\hat{c}_{t,b}$ are the annihilation operators of the two orbitals we retain, and $\hat{c}_m$ ($m\neq t,b$) annihilates one of the remaining orbitals.
We have introduced spinors $\phi^\dagger=(\hat{c}^\dagger_t,\hat{c}^\dagger_b)$ for 
the selected orbitals and $\chi^\dagger$ for remaining orbitals.
Integrating out the $(\chi^\dagger,\chi)$ fields, 
we obtain for the effective bilayer model at $\vec{d}$,
\be
\hat{h}_{eff}(\vec{k},\vec{d}) = \phi^\dagger\left[
{\bf h} +{\bf W}^\dagger\left(\mathcal{D}{\bf I} - {\bf H^R}\right)^{-1}{\bf W}\right]\phi.
\label{effective2l}
\ee
Here $\mathcal{D}$ is an energy near the center of the energy window over which our 
effective model is valid.  A natural choice is to set $\mathcal{D}$ to be the energy of VBM at $\vec{K}$ ($\vec{K}'$).
A detailed derivation can be found in Appendix \ref{pi}.

One feature of the effective model in Eq. \ref{effective2l} is that virtual hopping with the remaining orbitals is included in the perturbative correction term.  As a result no fitting is needed, and we are able to directly extract the layer-dependent potential $\Delta_{b,t}(\vec{d})$ and the inter-layer tunneling $\Omega_T(\vec{d})$, from which the crucial first-harmonic model parameters $(V_m,\psi,w_T)$ of the moir\'{e} continuum model can be obtained [\onlinecite{PhysRevLett.122.086402}]:
\be
\left[\hat{h}_{eff}(\vec{k},\vec{d})\right]_{ll}=
\Delta_l(\vec{d}) = 2V_m\sum_{1,3,5}\cos(\vec{G}_j\cdot\vec{d} - s_l\psi),
\ee
where $s_l=\pm 1$ for top ($l=t$) and bottom ($l=b$) layers respectively, and $\vec{G}_j$ are first-shell reciprocal lattice vectors. $V_m$ and $\psi$ are material-specific parameters determining the strength and the shape of the moir\'{e} potential, and $w_T = \Omega_T(\vec{d}=0)/3 = \vert \left[\hat{h}_{eff}(\vec{k},\vec{d}=0)\right]_{bt}\vert / 3$ is the tunneling strength.
The local approximation moir\'e band Hamiltonian is obtained  [\onlinecite{wu2019topological}] by letting $\vec{d} \to \theta \hat{z} \times \vec{r}$, corresponding to rigid rotation, and 
evaluating the $\vec{d}$-dependent terms in the Hamiltonian at $\vec{k}$ equal to a Brillouin-zone corner momentum.
Below, we use first-harmonic models to explore interaction effects, but higher-harmonic 
corrections can easily be added if more accurate material specific predictions are desired.   

\section{Results} 
\subsection{Hartree-Fock Theory}
Figs. \ref{fig:displacement}(b) and (c) illustrate the vertical structural 
relaxation of the metal positions in MoTe$_2$ and WSe$_2$ {\em vs.} stacking $\vec{d}$.
It can be seen that the metal-to-metal layer separation $Z$ is largest at AA stacking and 
shortest at MX and XM stackings, and that
the difference can be as large as 0.8 \r{A}, in good agreement with a recent large scale DFT study [\onlinecite{PhysRevB.109.205121}]. As discussed in Ref. [\onlinecite{PhysRevLett.118.147401}], the layer potential difference ($\Delta_t-\Delta_b$) is zero at AA stacking and largest in magnitude at MX and XM stackings. Consequently, the layer-separation variation significantly deepens the moir\'{e} potential. 
For the first-harmonic moir\'{e} model parameters, we obtain 
$(V_m,\psi,w_T)=(10.5{\rm meV},-89.948^\circ,11.1{\rm meV})$ for MoTe$_2$ and $(11.2{\rm meV}, -89.83^\circ,13.7{\rm meV})$ for WSe$_2$.
$V_m$ and $w_T$ are significantly larger in our approach than 
in previous estimates without vertical relaxation [\onlinecite{PhysRevLett.118.147401,PhysRevLett.121.026402}]. The increases of $V_m$ and $w_T$ are a consequence of local structural relaxation, as also observed in previous large-scale DFT calculations [\onlinecite{PhysRevLett.132.036501}]. 

\begin{figure*}
\begin{center}
\includegraphics[width=6.8in]{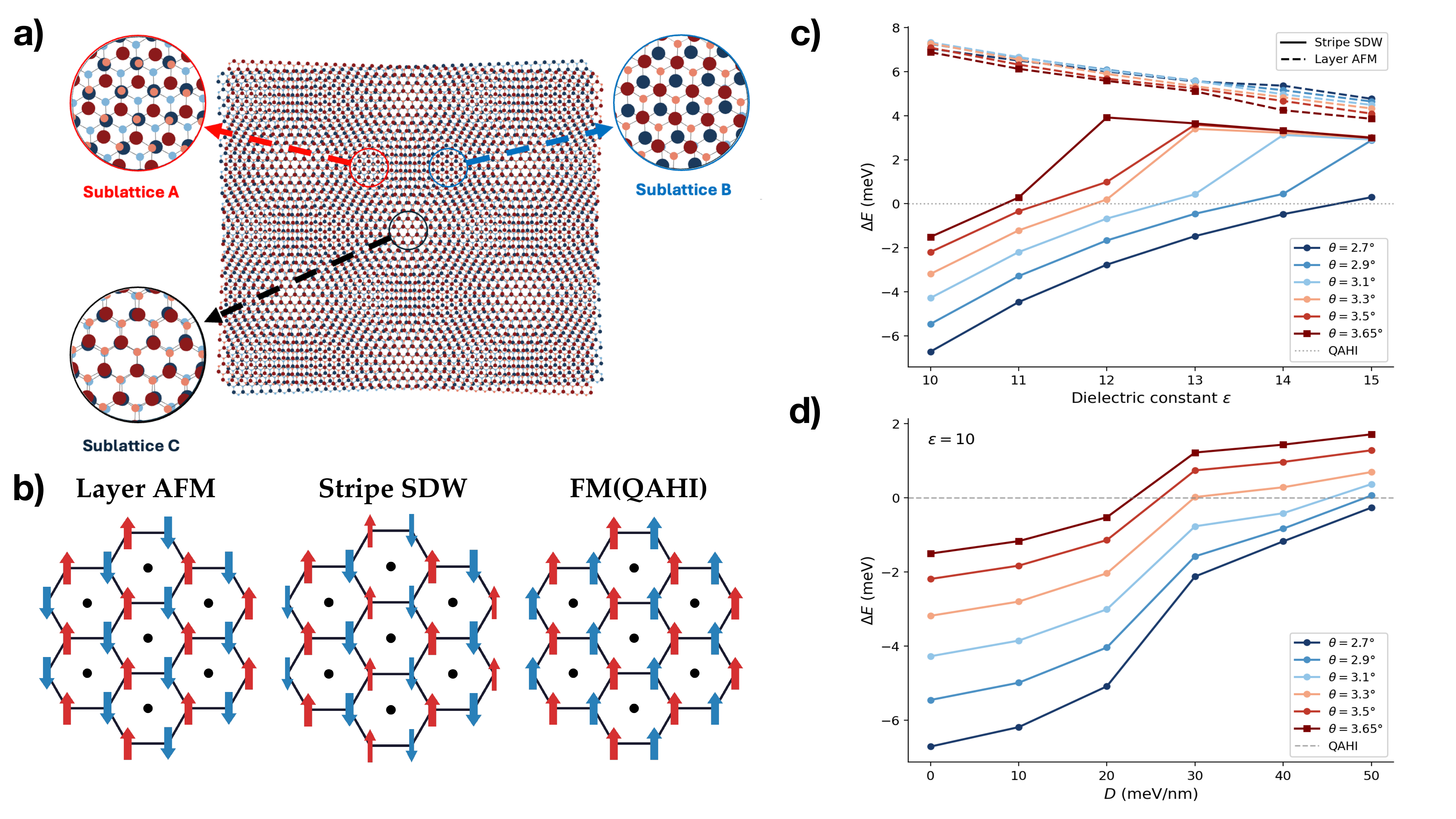}
\caption{\label{fig:WSe2-phase} (a) The three lowest energy bands of tMoTe$_2$ and 
tWSe$_2$ can be described by three band tight-binding models with the Wannier functions
defined in Ref. [\onlinecite{PhysRevX.13.041026}],
which are localized on sublattices A (red), B (blue), and C (black). 
The A and B Wannier wavefunctions are polarized to the top and bottom layers, respectively, while the sublattice C (black) Wannier wavefunction
is an equal weight linear combination of top and bottom layer components.
The A and B sublattices are more strongly weighted in the first two bands, and a two-band
model is often sufficient.  (b) Pictorial sketches of layer AFM, stripe SDW, and ferromagnetic (QAHI) states, 
described in terms of two-band models with A sublattice states in 
one layer and B sublattice states 
in the other layer. (c) Mean-field energies of the Layer AFM and stripe SDW states relative to the QAHI state 
($\Delta E \equiv \langle \hat{H}\rangle-\langle\hat{H}\rangle_{\rm QAHI}$) as a function of 
dielectric constant $\epsilon$ for $\theta\in[2.7^\circ, 3.65^\circ]$. 
(d) Same for the stripe SDW state as a function of displacement field $D$ at $\epsilon=10$.}
\end{center}
\end{figure*}

To explore competition between spin and charge orders, we employ the Hartree-Fock (HF) approximation, which accounts for important non-local exchange effects.
The phase diagram has been studied previously
by Qiu {\it et al.} [\onlinecite{PhysRevX.13.041026}] for the MoTe$_2$ case.
The details of our HF theory calculations
can be found in the Appendix \ref{hft}. We focus on the order parameters:
\beq
N&\equiv& \sum_{l=t,b}\sum_{\sigma=\uparrow,\downarrow} s_l n_{l,\sigma},\nn\\
M_F&=&\frac{1}{2}\sum_{l=t,b} \left( n_{l,\uparrow} - n_{l,\downarrow} \right),\nn\\
M^l_{AFM}&=&\frac{1}{2}\sum_{l=t,b}s_l\left( n_{l,\uparrow} - n_{l,\downarrow}\right).
\eeq
Here $N$ is the order parameter for 
vertically oriented ferroelectricity (layer polarization), 
$M_F$ for ferromagnetism,
and $M^l_{AFM}$
for layer antiferromagnetism.
For MoTe$_2$, we obtain results that are similar to those 
in Ref.~\onlinecite{PhysRevX.13.041026} with a phase 
diagram that includes a quantum anomalous Hall insulator (QAHI) FM state with a 
non-zero Chern number at $\theta=3.7^\circ$, in agreement with experiment.  
Here, we focus on features unique to WSe$_2$ at hole filling $\nu=1$.

\subsection{Competing orders in WSe$_2$ at hole filling $\nu=1$}

\begin{figure*}
\begin{center}
\includegraphics[width=6in]{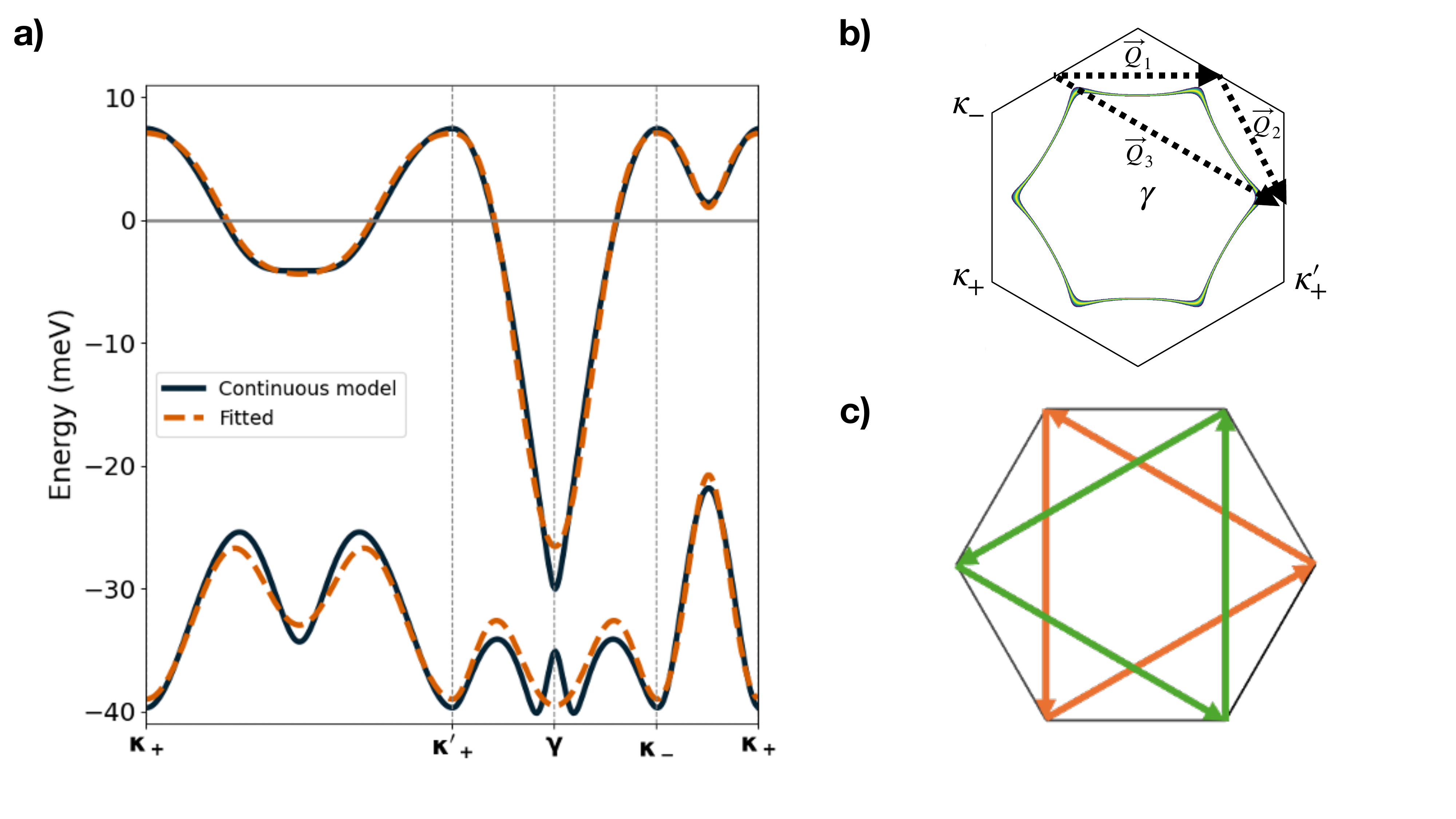}
\caption{\label{fig:FS} (a) Band structure of WSe$_2$ at $\theta=3.65^\circ$ based on continuum and approximate tight binding models. The fitted hopping parameters are $
(t_1,t_2,t_3,t_4,t_5,t_6) = (5.46,\,4.93,\,2.18,\,-0.56,\,-0.74,\,0.26)\,\text{meV}$ and the phase magnitude for $t_2$ is $\phi=0.65\pi$. (b) The Fermi surface has a high-density of states where it is closest to $M$-points (centers of the Brillouin-zone edges).
These favor SDW states at ordering wavevectors near $\vec{Q}_1=\vec{b}_1/2$,  $\vec{Q}_2=\vec{b}_2/2$ and $\vec{Q}_3=\vec{Q}_1+\vec{Q}_2$. (c) The arrow directions represent the positive phase winding of the complex $t_2$ hopping amplitude for $\uparrow$ spins. For $\downarrow$ spins, the phase winds in the opposite direction.}
\end{center}
\end{figure*}

The key difference between MoTe$_2$ and WSe$_2$ is in their 
continuum model effective masses ($m^*=0.62 m_{e}$ for MoTe$_2$ and 
$m^*=0.43 m_{e}$ for WSe$_2$, where $m_e$ is the bare electron mass).
In the continuum model, a smaller $m^*$ leads to more dispersive moir\'{e} bands 
and increased band separations, and this difference in band properties is likely the main reason why the phase diagrams are very different in MoTe$_2$ and WSe$_2$.
Although the WSe$_2$ bands are broader and more widely separated, band
velocities still vanish by symmetry at M-points and produce van Hove singularities
that have clear experimental fingerprints [\onlinecite{Xia2024}] when they coincide with the 
Fermi surface.  Motivated by the added significance of M-points in WSe$_2$, we allow the 
$(2,0)$ broken translational symmetry they favor and  
find the stripe spin-density wave (SDW) states illustrated in
Fig.~\ref{fig:WSe2-phase}(b) and Fig.~\ref{fig:FS}(b).  
The stripe state is described by the order parameter,
\be
M^s_{SDW}=\frac{1}{2N_k}\sum_{\vec{k}}^\prime\sum_{l=t,b}\sum_{\sigma=\pm} \sigma\langle c^\dagger_{\vec{k}+\vec{Q}_2,l,\sigma}c_{\vec{k},l,\sigma}\rangle,
\ee
where $N_k$ is the total number of $\vec{k}$ points in the moir\'{e} BZ, and the prime in
$\sum_{\vec{k}}^\prime$ recognizes that the BZ size is reduced
by half in the stripe state. The stripe SDW state is a mixture of the charge stripe order and layer antiferromagnetism with non-zero $M^l_{AFM}$ and $M^s_{SDW}$.

We have performed HF calculations that 
allow $N$, $M_F$, $M^l_{AFM}$, and $M^s_{SDW}$ order parameters
for the twist angles $\theta=2.7^\circ$ and, $\theta=3.65^\circ$, which have been experimentally investigated [\onlinecite{Xia2024, Knuppel2025}]. 
In the case of zero displacement field $D$, we did not find metastable states with ferroelectric $N \ne 0$ order over the range of model parameters considered. 
The FM state, which is stable over a broad range of filling factors in the 
MoTe$_2$ case, is always insulating and topologically non-trivial and is referred 
to below as the quantum anomalous Hall insulator (QAHI).
The ordered states that compete with the QAHI at zero displacement field are
a metallic layer AFM and a related insulating stripe SDW state discussed below.

We plot the mean-field energies of these states relative to that of the QAHI in Fig.~\ref{fig:WSe2-phase}
where we see that the stripe SDW state becomes the lowest energy state for 
$\epsilon < 14$ at $\theta=2.7^\circ$ and for $\epsilon<11$ at $\theta=3.65^\circ$.
In our HF calculations, the parameter $\epsilon$ is adjusted to be in the range of $10\leq\epsilon\leq 15$ partially 
account for screening beyond that provided by the hBN dielectric 
environment, so that these results provide a qualitative indication of ground state 
order trends. Based on Fig.~\ref{fig:WSe2-phase} and similar calculations for tMoTe$_2$, we conclude that stripe SDW state order is likely at small displacement 
fields in tWSe$_2$ but not in tMoTe$_2$ - and more likely at smaller twist angles. 
The layer AFM state can be viewed as a state with spontaneous layer polarizations 
of opposite senses for up and down spins.  Although it is metastable over a wide range of 
parameters, it is never the ground state at the examined twist angles.
Instead, each spin component has charge-density-wave instabilities at the M-point
wave vector, yielding the stripe SDW - the state that competes most effectively
with the QAHI state.  In our HF calculations, the stripe state is the
ground state at small $\epsilon$ (strong interactions) and is  
insulating at zero displacement field over the dielectric constant 
range we have studied ($10\leq \epsilon \leq 15$).
The physics changes qualitatively when a large displacement field is 
applied because the moir\'e minibands wavefunctions are strongly polarized to 
one layer and are no longer topologically non-trivial.
Since the sites available in one layer form a triangular lattice 
instead of a honeycomb lattice, insulating states at large displacement fields 
are likely to be the 3-sublattice antiferromagnetic states of conventional
Heisenberg models. Below, we discuss the superconducting states that compete 
with stripe SDWs, which we believe are likely to differ qualitatively from those 
that compete with conventional antiferromagnets.

\section{Proposed mechanism for superconductivity in twisted WSe$_2$} 

Now, we investigate the superconductivity observed at $\theta = 3.65^\circ$ and hole filling $\nu = 1$ at zero displacement field.  Experimentally, the superconducting phase is proximate to an insulating state that we interpret, based on our HF calculations, as a stripe SDW state.  We propose a mechanism for superconductivity based on
fluctuations of the order present in this state.

Fig.~\ref{fig:FS}(a) shows an accurate two-band honeycomb lattice model fit to the continuum description of WSe$_2$.  Note that sublattices A and B 
are associated with the top and bottom layers, respectively. 
Fits that allow up to sixth-neighbor hopping and allow 
the second neighbor hopping $t_2$ to be complex [\onlinecite{PhysRevLett.122.086402, Devakul_2021}] accurately reproduce the Fermi surface (Fig.~\ref{fig:FS}(b))
and the topological properties of the uppermost 
valence band. The phase structure of $t_2$ is crucial in obtaining the correct Chern numbers of the bands, and more remote hoppings are necessary to obtain accurate fits
across the Brillouin-zone. We note that the $\Gamma$ point of the second band shows a qualitative discrepancy with the continuum model, which arises because at twist angles near $\theta = 2.7^\circ$ [\onlinecite{Crepel3Sub}] the second and third bands touch at $\Gamma$, an effect that lies beyond our two-band description. This discrepancy is irrelevant for our purposes, as the superconductivity we study is driven by the uppermost valence band.

We model beyond mean-field fluctuations based on an interpretation of the honeycomb-lattice tight binding model.  The onsite Hubbard repulsion on sublattices A and B $U \approx e^2 / (\epsilon a_W)$, where $a_W$ is the spatial extent of the Wannier orbitals, from which 
we estimate that $U = 35\text{meV}$ for tWSe$_2$ using the same approximation.  The pattern of spin-ordering in the stripe AFM state places up spins and down spins on the top and bottom layer sites respectively of alternate zig-zag chains, as illustrated in Fig.~\ref{fig:WSe2-phase}.  Occupied nearest-neighbor sites always have opposite spins,
consistent with the superexchange interactions expected in the large $U/t_1$ limit [\onlinecite{layerpol}].
Superexchange between more remote neighbors is also expected to contribute to the 
fluctuation Hamiltonian, although we anticipate competing direct exchange interactions 
in the case of 2nd neighbors because they share the same layer.
These considerations suggest the effective Hamiltonian:
\beq
\hat{H}&=&\hat{H}_0 + \hat{H}_{J_1} + \hat{H}_{J_2} +  \hat{H}_{U}\nn\\
\hat{H}_{J_1}&=& J_1\sum_{<i,j>}\left( \vec{S}_i\cdot \vec{S}_j-\frac{1}{4}n_in_j    \right) \\
\hat{H}_{J_2}&=& J_2\sum_{<<i,j>>}\left( \vec{S}_i\cdot \vec{S}_j-\frac{1}{4}n_in_j    \right) \\
\hat{H}_U&=&U\sum_{i} n_{i\uparrow}n_{i\downarrow}
\eeq
\begin{figure}
\begin{center}
\includegraphics[width=3in]{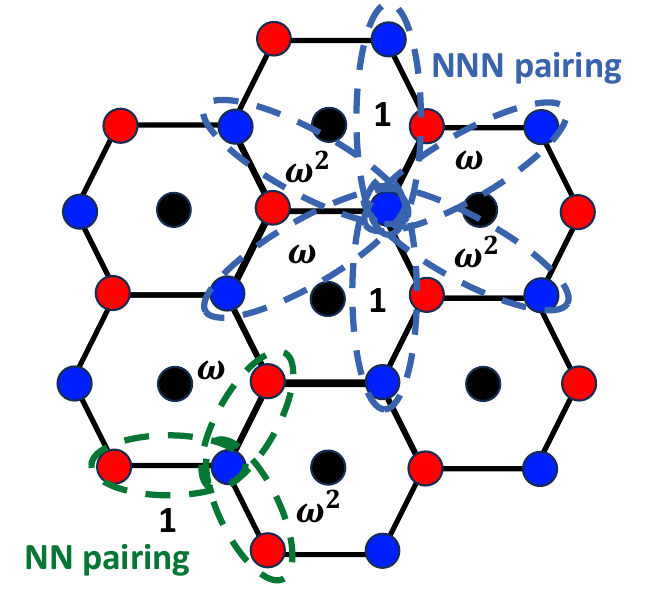}
\caption{\label{fig:pairing} Schematic illustration of pairing amplitude $\Delta_{ij}$ 
phases along different bonds in the nearest-neighbor (NN) and the next-nearest-neighbor (NNN) chiral pairing states ($\omega = \exp(i2\pi/3)$).} 
\end{center}
\end{figure}

The superconducting order parameters are defined as follows. $\Delta_a = \avg{a_{i\downarrow}a_{i\uparrow}}$ and $\Delta_b = \avg{b_{i\downarrow}b_{i\uparrow}}$ refer to the onsite pairings on sublattice $A$ and $B$ respectively. The inter-layer pairing via the nearest-neighbor (NN) bond between site $i$ on the sublattice $A$ and site $j$ on the sublattice $B$, can be decomposed into singlet and triplet components as 
\be
\Delta^{1,(s,t)}_{ij} = \avg{b_{j\downarrow}a_{i\uparrow}\mp b_{j\uparrow}a_{i\downarrow}}
\ee
Finally, the intra-layer singlet and triplet pairings on the same sublattice via next-nearest-neighbor (NNN) bonds can be defined as:
\beq
\Delta^{2,a,(s,t)}_{ij} &=& \avg{a_{j\downarrow}a_{i\uparrow}\mp a_{j\uparrow}a_{i\downarrow}}, \nn\\
\Delta^{2,b,(s,t)}_{ij} &=& \avg{b_{j\downarrow}b_{i\uparrow}\mp b_{j\uparrow}b_{i\downarrow}}.
\eeq

Because the pairing kernel transforms under the $D_{6h}$ point group, the NN and NNN pairings are allowed to have three inequivalent bonds on the honeycomb lattice. In this work, we focus on three different types of solutions: the isotropic state ($A_{1g}$ representation) with all three bond pairings being equal, the inter-layer chiral state due to NN pairing, and the intra-layer chiral state due to NNN pairing. The two chiral states ($E_{2g}$ representation) have a phase difference of $2\pi/3$ between bonds as shown in Fig.~\ref{fig:pairing}, and in the literature are often called $d\pm id$ for their singlet component and  $p\pm ip$ for their triplet component [\onlinecite{did1,did2,did3,FWuSuperc}].

\begin{figure*}
\begin{center}
\includegraphics[width=6.5in]{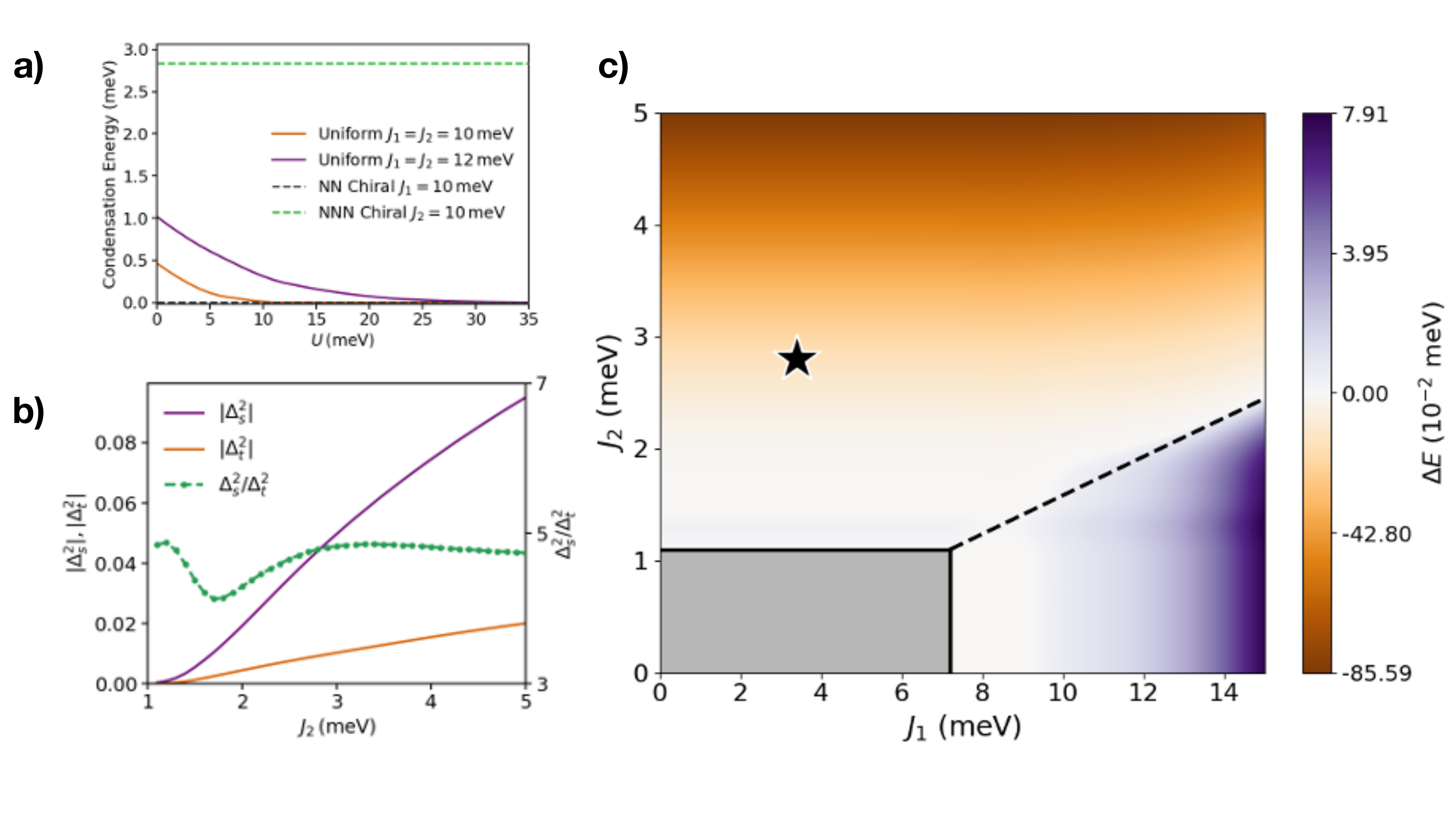}
\caption{\label{fig:SC} (a) Condensation energy per unit cell $= E_{S}-E_{N}$ for uniform and chiral representations is plotted against onsite repulsive interaction. (b) Magnitude of singlet and triplet order parameter of NNN pairing and their ratio {\it vs.} $J_2$. $\Delta^2_{s/t}=\avg{a_{j\downarrow}a_{i\uparrow}\mp a_{j\uparrow}a_{i\downarrow}}$. (c) The phase diagram for $U=35\,\mathrm{meV}$ characterized by the mean-field energy difference $\Delta E=E_{NN}-E_{NNN}$. The star corresponds to the values $J_1 = 3.4\,\mathrm{meV}$ and $J_2 = 2.78\,\mathrm{meV}$ estimated by $J_n=4t^2_n/U$. Grey area corresponds to the range of $(J_1,J_2)$ over which superconductivity was not found.}
\end{center}
\end{figure*}

With the order parameters defined above, the interaction terms can be decoupled in the particle–particle channel:
\beq
    \hat{H}_{J_1} &=& -\frac{J_1}{2}\sum_{<i,j>}\Delta_{ij}^{1,s}\left(a^\dagger_{i\uparrow}b^\dagger_{j\downarrow}-a^\dagger_{i\downarrow}b^\dagger_{j\uparrow} \right)+h.c+const \nn \\
    \hat{H}^a_{J_2} &=& -\frac{J_2}{2}\sum_{<<i,j>>} \Delta^{2,a,s}_{ij}\left(a^\dagger_{i\uparrow}a^\dagger_{j\downarrow}-a^\dagger_{i\downarrow}a^\dagger_{j\uparrow} \right)+h.c +const \nn \\
    \hat{H}^b_{J_2} &=& -\frac{J_2}{2}\sum_{<<i,j>>} \Delta^{2,b,s}_{ij}\left(b^\dagger_{i\uparrow}b^\dagger_{j\downarrow}-b^\dagger_{i\downarrow}b^\dagger_{j\uparrow} \right)+h.c +const \nn \\
    \hat{H}_{U} &=& U\sum_{i}\left(\Delta_{a}a^\dagger_{i\uparrow}a^\dagger_{i\downarrow}+\Delta_{b}b^\dagger_{i\uparrow}b^\dagger_{i\downarrow} \right) +h.c+const \nn
\label{mfd}
\eeq
We self-consistently calculate the superconducting order parameters at hole filling $\nu = 1$. We begin by analyzing the case with $U = 0$. 
We find that the two chiral states never coexist. 
In contrast, the isotropic $A_{1g}$ state induced by $J_1$ and $J_2$ can coexist and mutually enhance each other. This occurs because all isotropic states share the same translational and rotational symmetries as the crystal lattice; 
their coexistence does not violate any symmetry principles. However, this means that an effective onsite pairing will also appear in an isotropic state, whereas in both NN and NNN chiral states onsite pairing is strictly forbidden by symmetry.
As shown in Fig.~\ref{fig:SC}(a), when a strong repulsive interaction $U$ is included, the isotropic $A_{1g}$ states are strongly suppressed, but the chiral states are unaffected since they do not have onsite pairing.  When realistic onsite repulsion is included, chiral states become the dominant superconducting instability.
We therefore focus on the competition between NN and NNN chiral pairings as a function of $J_1$ and $J_2$. Fig.~\ref{fig:SC}(c) plots the phase diagram for tWSe$_2$ at the twist angle $\theta=3.65^\circ$ with $U=35\,\mathrm{meV}$. We observe that the NNN pairing channel generally dominates over the NN pairing channel; the NN pairing becomes the leading instability only when $J_1/J_2\gg 1$.

Similar behavior has been obtained and discussed in the recent work by Tuo {\it et al.} [\onlinecite{Tuo2025}], which can be understood as arising 
from the structure of the complex hopping term $t_2$ illustrated
in Fig.~\ref{fig:FS}(c).  The Cooper pairing is strong when the electronic states at $\vec{k}\uparrow$ and $-\vec{k}\downarrow$ have the same phase winding and layer projection. It can be seen that states of
 $a_{\vec{k}\uparrow}$, $b_{\vec{k}\downarrow}$, $a_{-\vec{k}\downarrow}$ and $b_{-\vec{k}\uparrow}$ will all have the same phase winding. Moreover, since the layer projection is invariant under time-reversal, states of $\vec{k}$ and $-\vec{k}$ 
 have the same layer polarization. 
 Therefore, the pairing between $(\vec{k}\uparrow)$ and $(-\vec{k}\downarrow)$ will be favored in the intra-layer (NNN) channels. 

Finally, we would like to remark on the mixture of singlet and triplet 
components discussed in Ref. [\onlinecite{Tuo2025}], which is always present 
on the basis of symmetry considerations. However, the degree of mixture could have a hint on the pairing mechanism. In Ref. [\onlinecite{Tuo2025}], effective NN and NNN attractive interactions are introduced as the source of pairing so that their mean-field decoupling contains a triplet part directly. As a result, their triplet component can become large in some parameter space. Our pairing mechanism originates from AFM spin fluctuations, and consequently our mean-field decoupling presented in Eq. \ref{mfd} contains only singlet components. As a result, the triplet component is much smaller in our scenario because it does not have any energetic benefit. We follow the definition given in Ref. [\onlinecite{Tuo2025}] to extract the triplet component, and we confirm that the triplet component is indeed much smaller as shown in Fig.~\ref{fig:SC}(b). We note that chiral mixed-parity d/p pairing has been identified across a range of theoretical approaches, including continuum [\onlinecite{FWuSuperc}] models, single-band [\onlinecite{zegrodnik2025}] and three-band [\onlinecite{Tuo2025,Fischer_Sup}] of WSe$_2$, predominantly at larger twist angles, consistent with our finding that this pairing symmetry is robust over a wide range of model parameters Fig.~\ref{fig:SC}(c).

\section{Summary and conclusion}
In this paper, we have presented an improved theoretical framework to include the effect of local relaxation along the $c$-axis into the modeling of moir\'{e} electronic structures. We have shown that our model has a significant improvement compared to the model without relaxation, and it is able to produce results close to those of
large-scale DFT with much lower computational cost. Moreover, all the TB models are directly obtained from first-principles calculations, and all the virtual hoppings via orbitals other than the selected ones are included exactly through a path-integral method. Consequently, no extra fitting is needed to construct the effective bilayer model. We would like to further emphasize that our approach can be employed to construct a very wide range of effective models, e.g., moir\'{e} models near $\Gamma$ points or moir\'{e} models for the conduction bands.  The number of orbitals in the final effective model can be more than two, depending on the systems under study. This framework will be very useful for studying various types of heterostructures of TMDs in the future.

We have constructed the continuum models for MoTe$_2$ and WSe$_2$ using our proposed method. While our model for MoTe$_2$ reproduces the phase diagram obtained in existing literature, we find that in WSe$_2$ layer AFM and stripe SDW states compete most closely with QAHI
states in terms of mean-field energy.  Our HF calculations show that the stripe SDW state is insulating and becomes the lowest energy state if the dielectric constant is in the range of $\epsilon < 14$ at twist angle $\theta = 2.7^\circ$, which could explain the observation of an insulating state with zero magnetization [\onlinecite{Knuppel2025}]. We have also shown that the layer AFM is a metallic state due to its preservation of $C_{2y}$ rotational symmetry, and we propose that superconductivity can emerge from this metallic state with an AFM spin fluctuation pairing mechanism. We investigate superconductivity emerging at hole filling $\nu=1$ in twisted bilayer WSe$_2$ and demonstrate that it is driven by antiferromagnetic exchange originating from strong electronic correlations. Within an effective honeycomb model in which sublattices A and B represent the top and bottom layers, we find that the large onsite repulsion suppresses any superconducting state containing onsite pairing components. Consequently, isotropic $A_{1g}$ states are strongly disfavored. In contrast, chiral superconductivity, which does not involve onsite pairing, remains robust against repulsion and becomes the dominant instability. We find that the leading superconducting channel is governed by intra-layer (next-nearest-neighbor) pairing. The resulting chiral state exhibits a mixture of singlet and triplet components, with the singlet part always remaining dominant throughout the physically relevant parameter range. Our results point toward a correlation-driven superconducting phase that naturally emerges from a correlated insulating state.

\subsection{Acknowledgement}
E.J. and S.O. contributed equally to this work.
This work was supported by the Air Force Office of Scientific Research
Multi-Disciplinary Research Initiative (MURI) entitled, “Cross-disciplinary
Electronic-ionic Research Enabling Biologically Realistic Autonomous Learning
(CEREBRAL)” under Award No. FA9550-18-1-0024 administered by Dr. Ali Sayir.
A.H.M. was supported by a Simons Foundation Targeted Grant under 
Award No. 896630.

\appendix
\section{Path-integral formalism}
\label{pi}
As shown in the main text, the TB model in the spherical harmonic orbitals $\hat{H}(\vec{k},\vec{d})$ can be separated into three parts: a $2\times 2$ matrix describing the two orbitals ($\phi$) to be kept, a $42\times 42$ matrix describing the remaining orbitals ($\chi$) to be
integrated out, and a $42\times 2$ matrix together with its Hermitian counterpart describing the coupling between these two groups of orbitals. The idea is to use the path integral to 'integrate' out other orbitals to create the effective two orbital models. Starting from the general formalism for the partition function in the field-theory approach:
\begin{widetext}
\beq
Z&=&\int D[\bar{\phi}]D[\phi]D[\bar{\chi}]D[\chi] e^{-\frac{S}{\hbar}},\nn\\
S&=&\int_0^\beta d\tau\left[\bar{\phi}\left(\partial_\tau{\bf I} + {\bf h}\right)\phi + \bar{\chi}\left(\partial_\tau{\bf I} + {\bf H^R}\right)\chi + \bar{\chi} {\bf W} \phi + h.c. \right]\nn\\
&=&\frac{1}{\beta}\sum_{i\omega_n} \left[\bar{\phi}(i\omega_n)\left(-i\omega_n{\bf I} + {\bf h}\right)\phi(i\omega_n) + \bar{\chi}(i\omega_n)\left(-i\omega_n{\bf I} + {\bf H^R}\right)\chi(i\omega_n) + \bar{\chi}(i\omega_n) {\bf W} \phi(i\omega_n) + h.c. \right]
\eeq
\end{widetext}
where $(\bar{\phi},\phi,\bar{\chi},\chi)$ are all complex Grassmann variables, and in the last step we have transformed to Matsubara frequency $i\omega_n$.
We can now integrate out $D[\bar{\chi}]D[\chi]$ to obtain 
\beq
Z&=&det\left(-i\omega_n{\bf I} + {\bf H^R}\right)\int D[\bar{\phi}]D[\phi] e^{-\frac{S_{eff}}{\hbar}},\nn\\
S_{eff}&=&\frac{1}{\beta}\sum_{i\omega_n} \bar{\phi}(i\omega_n)\left[-i\omega_n{\bf I} + {\bf h} - {\bf W}^\dagger{\bf G}^\chi(i\omega_n){\bf W}\right]\phi(i\omega_n)\nn\\
\eeq, where
\be
{\bf G}^\chi(i\omega_n)=\left(-i\omega_n{\bf I} + {\bf H^R}\right)^{-1}
\ee
is the Green's function (propagator) of the orbitals other than the selected two orbitals.
With this effective action, we can obtain the Green's function for $\phi$ and then use the analytical continuation to obtain the retarded Green's function
\be
{\bf G}^\phi(E) = \left[- E{\bf I} + {\bf h} +{\bf W}^\dagger\left(E{\bf I} - {\bf H^R}\right)^{-1}{\bf W}\right]^{-1}
\ee
As a result, we can find the effective Hamiltonian at energy $E$ as
\be
\hat{h}_{eff} = \phi^\dagger\left[
{\bf h} +{\bf W}^\dagger\left(E{\bf I} - {\bf H^R}\right)^{-1}{\bf W}\right]\phi
\ee

\section{Hartree-Fock theory}
\label{hft}
We consider the screened Coulomb interaction in the form of
\be
V_{l,l'}(\vec{q})= \frac{2\pi e^2}{\epsilon q}\tanh(qd_G)\left[1+(1-\delta_{l,l'})(e^{-qd}-1)\right],
\label{screenedvc}
\ee
where $d_G$ is the distance from the gate to sample, and $d$ is the layer distance. In general, $d_G \gg d$, and we will use the relaxed $d$ at AA stacking, which is 7.7 \AA for MoTe$_2$ and 7.1 \AA for WSe$_2$.

We adopt the Hartree-Fock self-energies derived in Ref. [\onlinecite{PhysRevLett.124.097601}] in our mean-field calculations, and we add two more mean-field self-energies.

{\bf Inter-valley coherence} -- we consider the inter-valley coherence by allowing the following mean-field self-energy:
\begin{widetext}
\be
\left[{\mathcal H^{\tau,\tau'}_F(\vec{k})}\right]_{\alpha,\vec{G};\beta\vec{G}'}=-\frac{1}{A}\sum_{\vec{G}'',\vec{k}'}V(\vec{G}''+\vec{k}'-\vec{k})\rho_{\tau,\alpha,\vec{G}+\vec{G}'';\tau',\beta,\vec{G}'+\vec{G}''}(\vec{k}'), \tau\neq\tau',
\ee
\end{widetext}
where 
\be
\rho_{\tau,\alpha,\vec{G}+\vec{G}'';\tau',\beta,\vec{G}'+\vec{G}''}(\vec{k}') = \langle \hat{c}^\dagger_{\tau',\beta,\vec{G}'+\vec{G}''}\hat{c}_{\tau,\alpha,\vec{G}+\vec{G}''}\rangle_{MF}
\ee
is the density matrix evaluated in the mean-field ground state $\langle\cdots\rangle_{MF}$.

{\bf Inter-layer Hartree potential} -- 
A typical argument to ignore the Hartree term with $\vec{G}-\vec{G}'=0$ is the background charges from the ions neglected in the Hamiltonian. However, since we consider the possibility of the  inter-layer ferroelectricity (FE), the energy cost due to the charge imbalance between layers should be included. However, this energy cost is in fact from the Hartree term with zero momentum. As a result, we need to include the part that can penalize the charge imbalance between layers. We note that the same term has been included, as discussed in Ref. [\onlinecite{PhysRevLett.131.046402}].

First of all, the interaction kernel at $\vec{q}=0$ is finite if using the screened Coulomb interaction:
\be
V(\vec{q})=\frac{2\pi e^2}{\epsilon q}\tanh(qZ)\,\,\,\to\lim_{q\to 0}\to\,\,\,
V(0) = \frac{2\pi e^2 Z}{\epsilon},
\ee
At $\vec{q}=0$, the bare interaction Hamiltonian is in fact just
\be
\frac{V(0)}{2A}\left(\hat{N}_b+\hat{N}_t\right)\left(\hat{N}_b+\hat{N}_t\right)
\ee
where
\be
\hat{N}_\alpha = \sum_{\vec{p}} \sum_{\tau} \hat{c}^\dagger_{\vec{p},\tau,\alpha} \hat{c}_{\vec{p},\tau,\alpha}
\ee
is the total electron number operator for layer $\alpha$.
The term of $\hat{N}_b \hat{N}_t$ will lead to the exchange terms favoring FE which are already included in the Fock terms. As a result, the terms we should keep to penalize the layer charge imbalance are
\be
\frac{V(0)}{2A}\left(\hat{N}_b\hat{N}_b + \hat{N}_t \hat{N}_t\right),
\ee
and then we can apply the mean-field theory to obtain
\be
\Sigma_{\alpha}^{H, MF}(0) = V(0)\left(\rho_b\hat{N}_b + \rho_t \hat{N}_t\right) + constants,
\ee
where
\be
\rho_\alpha = \frac{1}{\Omega_k}\sum_{\vec{k}}\sum_{\vec{G}} \sum_{\tau} \langle\hat{c}^\dagger_{\vec{k},\tau,\alpha,\vec{G}} \hat{c}_{\vec{k},\tau,\alpha,\vec{G}}\rangle,
\ee
is the total charge density in layer $\alpha$. $\Omega_k$ is the number of $\vec{k}$ points used in the calculation, and $A$ is the area of the moir\'{e} unit cell. We can define
\be
\rho_{tot} = \rho_b+\rho_t\,\,\,,\,\,\,
\Delta \rho = \rho_t-\rho_b,
\ee
and then we have
\be
\Sigma_{\alpha}^{H, MF}(0) = \frac{V(0)}{2A}\left[\rho_{tot}\left(\hat{N}_b + \hat{N}_t\right) + \Delta\rho\left(\hat{N}_t - \hat{N}_b\right) \right] + constants,
\ee
The first term is just a constant shift that can be absorbed into the chemical potential $\mu$. Therefore, the self-energy term needed is
\be
\Sigma_{\alpha}^{H, MF}(0) = \frac{V(0)\Delta\rho}{2A}\left(\hat{N}_t - \hat{N}_b\right) + constants,
\ee
 
\bibliography{DFT,TMD}{}

\begin{thebibliography}{10}
\bibitem{layerpol}
In our mean-field calculations spontaneous layer polarization of like spins,
  favored by direct exchange interactions, also plays a role in stabilizing the
  SDW states.
\end{thebibliography}

\end{document}